# Accurate real time localization tracking in a clinical environment using Bluetooth Low Energy and deep learning

Zohaib Iqbal, Da Luo, Peter Henry, Samaneh Kazemifar, Timothy Rozario, Yulong Yan, Kenneth Westover, Weiguo Lu, Dan Nguyen, Troy Long, Jing Wang, Hak Choy, Steve Jiang*

Medical Artificial Intelligence and Automation Laboratory, Department of Radiation Oncology, University of Texas Southwestern Medical Center, Dallas, TX, United States of America

* Steve.Jiang@utsouthwestern.edu

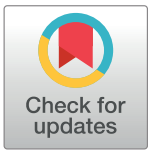

## Abstract

Deep learning has started to revolutionize several different industries, and the applications of these methods in medicine are now becoming more commonplace. This study focuses on investigating the feasibility of tracking patients and clinical staff wearing Bluetooth Low Energy (BLE) tags in a radiation oncology clinic using artificial neural networks (ANNs) and convolutional neural networks (CNNs). The performance of these networks was compared to relative received signal strength indicator (RSSI) thresholding and triangulation. By utilizing temporal information, a combined CNN+ANN network was capable of correctly identifying the location of the BLE tag with an accuracy of 99.9%. It outperformed a CNN model (accuracy = 94%), a thresholding model employing majority voting (accuracy = 95%), and a triangulation classifier utilizing majority voting (accuracy = 95%). Future studies will seek to deploy this affordable real time location system in hospitals to improve clinical workflow, efficiency, and patient safety.

**Data Availability Statement:** All of the images used for the research are uploaded to the Github repository: https://github.com/zoball/BLE-Tracking-with-Deep-Learning. These images comprise the entire data set (more than the minimal data set shown in the paper). All of the images are in .png format and are in labelled folders, which makes the data easy to read and classify based on directory location.

**Funding:** The authors received no specific funding for this work.

## Introduction

In recent decades, the field of medicine has evolved at an alarming rate. The advent of innovative technologies has fueled the progression of cutting-edge treatment and quality patient care. In order to carry on this evolution, technological advancements continue to be integrated and adopted into the clinic. Wearable sensors [1–3] provide a new avenue for ensuring patient health by providing physicians with tools to monitor several biochemical metrics, including blood pressure, oxygen saturation, heart rate, and several others, outside of the treatment facility. Furthermore, with the proliferation of wireless technology, a patient's necessary biochemical information can be transmitted directly to a clinic from almost anywhere in the world. However, it is infeasible for physicians or other medical staff to constantly view these biometrics in real time to detect any abnormalities in a patient's condition.

Fortunately, continuous monitoring is possible by utilizing deep learning methods. With proper training, supervised learning techniques such as convolutional neural networks [4, 5] are capable of analyzing large data in order to yield meaningful results. First, a training data set is collected and properly labelled. This data set is then used to train a neural network model to perform the desired task. Finally, this model is tested by employing an independent testing data set which the network has never encountered. By monitoring biometric and workflow data, a smart clinic consisting of various deep learning models may play a large role in improving the efficiency and safety of a clinic, especially in the setting of a radiation oncology facility where quality assurance and patient safety are a primary concern [6].

In order to monitor clinical workflow, it is essential to develop an accurate real time location system (RTLS) to track patients and clinical staff. Currently, the main-stream commercial method that is used for indoor tracking is radio-frequency identification (RFID) technology [7, 8]. Unfortunately, the proprietary hardware required for this system is expensive and requires extensive cabling. On the other hand, Bluetooth Low Energy (BLE) technology provides an alternative method for indoor tracking [9–15]. Due to the accessibility of BLE tags and detectors, this technology is versatile and can be used in combination with various hardware. However, affordable implementations of this method may have difficulties locating BLE tags with the conventional received signal strength intensity (RSSI) triangulation method. This is primarily due to the fact that the RSSI values are unstable because of higher noise in the BLE tag signal, thus producing localization results that may be inaccurate.

This study focuses on investigating the feasibility of an RTLS system that utilizes deep learning for tracking wearable tags assigned to patients and clinical staff. Using BLE technology, a sensor network was installed throughout our radiation oncology clinic for indoor tracking. Positional data were collected in several different areas using Bluetooth tags, and these data were classified using novel deep learning methods, as well as more traditional techniques such as RSSI thresholding and triangulation.

## Materials and methods

### Hardware

The Bluetooth sensor network was built with two key factors in mind: 1) ability to accurately detect the general position of a BLE tag and 2) affordable cost to allow for versatile changes and easy adoption into other clinics. A large number of Raspberry Pis (RPI 3 Model B with Raspbian kernel 4.4.38-v7) equipped with Cirago BTA8000 USB dongles were installed throughout the three floors of the radiation oncology clinic at the University of Texas Southwestern Medical Center. The installation was performed with the guidance and permission of the institution. The majority of the Pis were placed above the ceiling tiles of the first and second floors, where patient traffic is high. In total, 120 Pis were installed on the first floor, 24 Pis were installed on the second floor, and 9 Pis were installed on the third floor. These Pis were connected to power sources and were communicating directly to a MongoDB database. Only a small portion of the first floor was used for the experiments conducted in this study, and this limited region can be seen in Fig 1A. The Pis, henceforth referred to as nodes, are represented by red squares in the schematic diagram and are present in all exam rooms and select hallways.

The RadBeacon Dot (Radius Networks) BLE tag was used for all experiments in this study. This tag was powered by a replaceable battery and was functioning with a power of +3 dB and sending signals at a rate of 10Hz. Before collecting a large data set, two RadBeacon Dot tags were evaluated in two separate exam rooms. The RSSI of the two tags were recorded as a function of the distance from the nodes, and the results can be seen in Fig 2. It is clear that the signal strength drops as the tag is further from the node, which is desirable when detecting the tag's location.

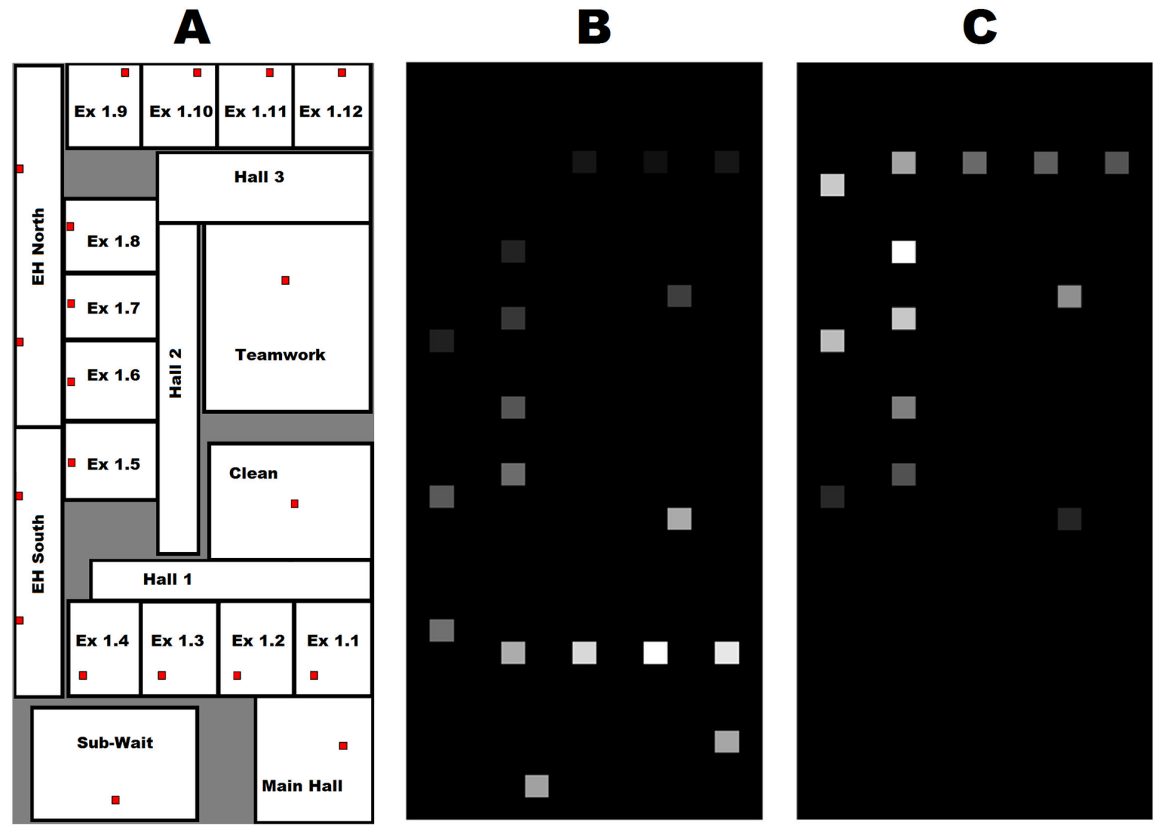

**Fig 1. Spatial representation of data.** A) A schematic of the floor plan for the first floor exam area where the experiments were conducted; B) A representative grayscale image used for training when the tag is in exam room 1.2, and C) A representative grayscale image when the tag is in exam room 1.8. In the grayscale images, each pixel with signal corresponds to a Raspberry Pi node, and therefore there are only 20 elements in the 34 x 15 image that vary signal as a function of time and tag position. White represents a high RSSI value and black represents a low RSSI value in the images.

https://doi.org/10.1371/journal.pone.0205392.g001

### Data collection

A total of ten tags were utilized to gather data from the twenty-one zones of interest using a test setup. It is important to note that no patients were involved in the data collection process. A researcher would equip a tag either at the waist or chest area. Then, the researcher would enter a zone of interest and move randomly throughout that zone for a total time of four minutes. The twenty-one zones, which are outlined with black border lines in Fig 1A, include the exam rooms (1.1 – 1.12), elevator halls north and south, corridor halls (1–3), main hall, sub-wait area, clean room, and teamwork room. The data collected during this process were the RSSI values at each node, and these values varied from -150 to -60. Therefore, twenty-one RSSI values were recorded at a rate of 10Hz for the entire four minute duration. The researcher would then move to the next zone, and repeat this process until data collection for all tags and zones were complete. For this experiment, the ten tags were in the same room at the same time for data collection to replicate a clinical scenario where multiple staff members and patients occupy the same space, for example the Teamwork room. However, there was no signal interference due to the presence of multiple tags in the same room, which was validated in a separate experiment not shown here. After data collection, the RSSI data for each tag could be accessed for each zone using the corresponding time stamp and label in the database.

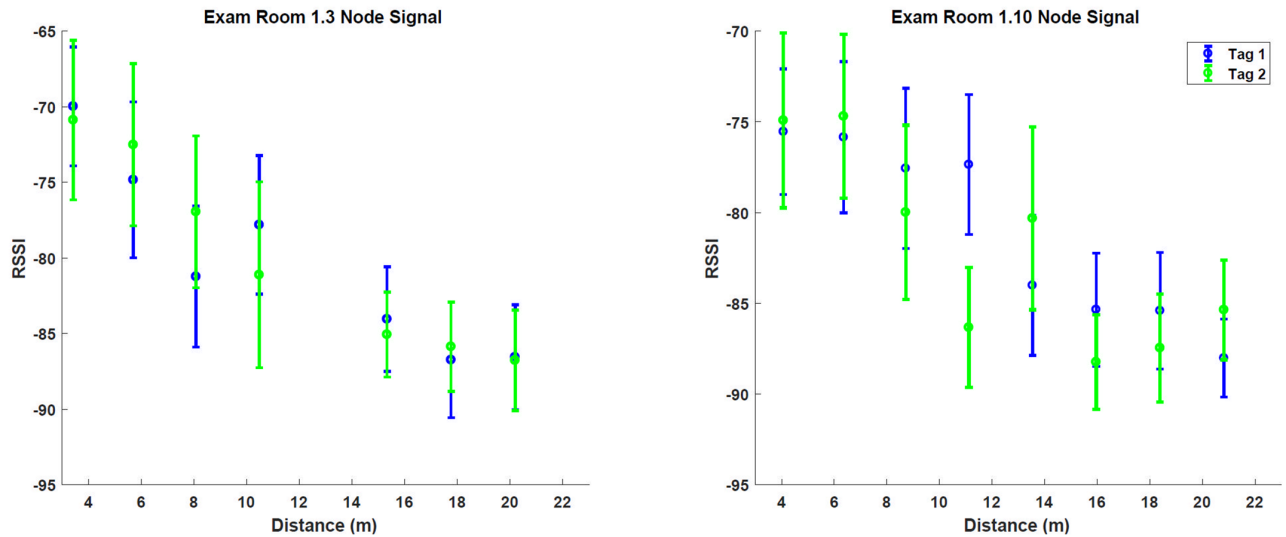

**Fig 2. Signal intensity as a function of distance from node.** The relative received signal strength (RSSI) is shown as a function of distance from the node. Two RadBeacon Dot (Radius Networks) tags were tested in two different exam rooms, and both show decreasing signal strength as the tags are placed further from the Raspberry Pis.

https://doi.org/10.1371/journal.pone.0205392.g002

## Data processing

The data processing focused on converting the data into a form that a convolutional neural network could easily read and evaluate. The first processing step averaged the data to effectively reduce the data acquisition rate from 10Hz to 1Hz. While this averaging reduced the amount of data by a factor of ten, this process helped subdue random noise which could potentially cause errors in classification. From a mathematical standpoint, the original data for each second could be represented as a 21 x 10 matrix, where 21 is the total number of nodes receiving signal, and 10 is the number of samples. After the averaging process, each second of data was represented by a vector of length 21. Next, the RSSI signals were thresholded so that any signals below −100 RSSI would be set to zero. This thresholding was performed for two primary reasons. First, values below this number corresponded to node locations that were very far from the tag location, and therefore would provide minimal additional information for localization purposes. Second, signals below this value may be unreadable and would give non-numeric values for evaluation. Then, all of the remaining elements in the signal vector were normalized so that the maximum would be one and the minimum would be zero:

$$E_{norm} = \frac{E_{sig} - V_{min}}{V_{max} - V_{min}} \quad (1)$$

In Eq 1, $E_{norm}$ is the normalized element, $E_{sig}$ is the raw RSSI value of the corresponding element, $V_{min}$ is the minimum RSSI value in the signal vector, and $V_{max}$ is the maximum RSSI value in the signal vector.

Finally, these normalized signals were mapped onto a 34 x 15 pixel grayscale image, as shown in Fig 1B and 1C. This 34 x 15 pixel map is representative of the space shown in Fig 1A, and each pixel corresponds to a specific node's physical location. Since there are only 21 nodes, a maximum of 21 pixels will have signal intensity at fixed locations. The normalized signals for each second were mapped onto their physical node locations for this graphical depiction. Each image represented one second, and therefore this process generated 240 grayscale images for each tag in each zone. White indicates that the node was receiving very high signal at a particular time point. Conversely, black indicates that a node was receiving very little signal at that time. As the tag is tracked during movement and in different zones, the signal properties of the map change drastically. Finally, these images were saved according to tag, zone, and time for training and testing purposes.

## Neural network architecture and training

All neural networks were written and evaluated in Python 3 using both the Tensorflow [16] and Keras [17] packages. The first network design is shown in Fig 3. The input layer is the 34 x 15 pixel grayscale image saved from the data processing step. First, a 7x7 convolution layer with a rectified linear unit (ReLU) [18] activation function was used to create 32 feature maps. Convolution of the image allows for important spatial relationships to be identified and extracted. A 7x7 kernel was used so that neighboring nodes would be included into the convolution process, ensuring that the network would learn the relationships between these nodes. The convolutional layer was followed by maximum pooling, reducing the size of the feature maps by a factor of two in each dimension. Then a 3x3 convolution layer with the ReLU activation function and another max pooling layer were also used. The last 32 feature maps were flattened and were fully connected to 128 neural nodes. After flattening, it is possible to use dense layers for classification purposes, which is common practice in deep learning. Finally, a layer with softmax activation was employed to yield probabilities for each zone. For this CNN, the maximum probability in this vector was used to determine the classification of the image.

The above CNN was trained using 70% of the collected data and was tested using 30% of the data. This translated into using data from seven tags in each zone for training, and three tags in each zone for testing. The training data is sent through the CNN and an associated error is calculated based on the model's classification results. Based on the error, the model refines the weights, and this process is repeated until all of the training data are used. Testing data is applied in the same manner, however the weights are not updated, and therefore the model is unchanged after evaluating testing data. Dropout [19] was set to 0.3 for the convolutional layers and for the fully connected layer in order to prevent over-fitting in the model.

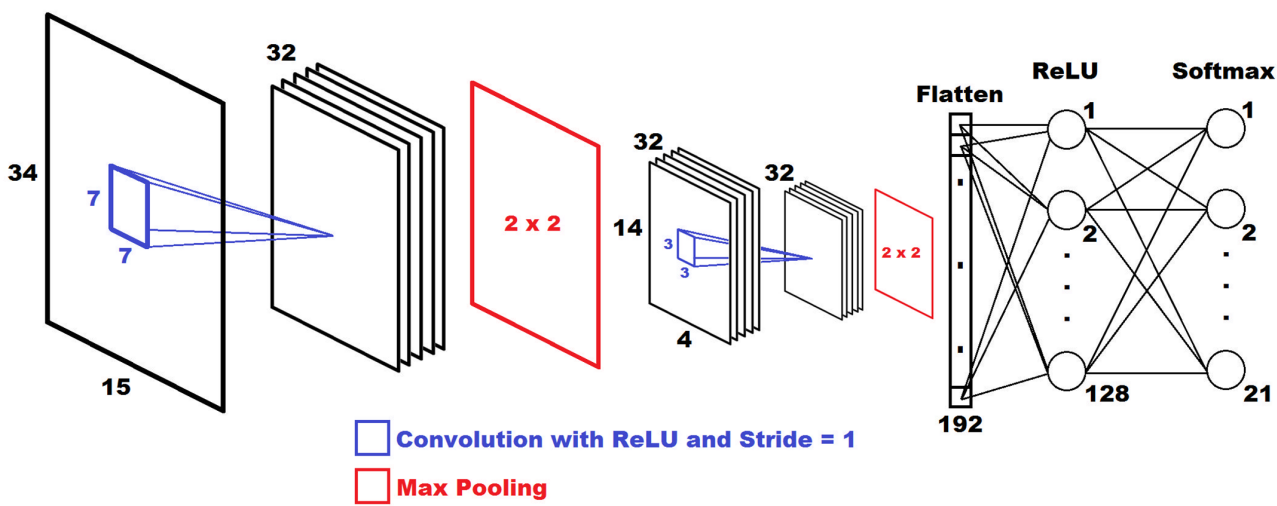

**Fig 3. Convolutional neural network architecture.** The structure of the convolutional neural network is displayed. Initially, two convolution and max pooling layers are used. Then, the feature maps are flattened and input into a fully connected network. Finally, a layer with the softmax activation function is used to obtain the classification.

https://doi.org/10.1371/journal.pone.0205392.g003

The loss function was set to minimize the categorical cross-entropy using the Adam optimizer [20]. The training was performed with a batch size of 44 and 85 epochs.

To take advantage of the temporal information of each tag, a separate artificial neural network (ANN) was also designed, and can be seen in Fig 4. This artificial network required thirty consecutive seconds of data from a particular tag in order to make a prediction. First, the trained CNN above would classify each individual image to yield a probability vector. The probability vectors for all thirty seconds would then be combined and used as the input layer for the ANN. This included the current time, $t$, as well as the twenty-nine previous seconds ($t - 1, t - 2, \ldots, t - 29$). The ANN was designed with a total of three layers. The first two layers each include 125 neural nodes activated using ReLU. The final layer is a softmax layer yielding the final probability vector, the maximum of which is the final classification.

Similar to the training for the CNN, this ANN was trained using 70% of data in all of the training zones, and was tested using the remaining 30% of data. Since the first twenty-nine seconds of data recorded do not have an ample temporal history, they were not used in the training or testing of the ANN. The rest of the data were used by implementing a sliding window for data grouping. The size of the window was thirty seconds and the window was shifted by one second for the next set of data. In this manner, the data size was preserved while also providing thirty consecutive seconds for training and testing purposes. In addition, this procedure was repeated for varying history lengths in order to determine the affect of the temporal information on accuracy. However, a thirty second history was used for the final comparisons. The first two layers were trained with a Dropout of 0.5 to prevent over-fitting. Once again, the categorical cross-entropy was minimized using the Adam optimizer to train the network. A batch size of 25 was chosen and the training was performed using 15 epochs.

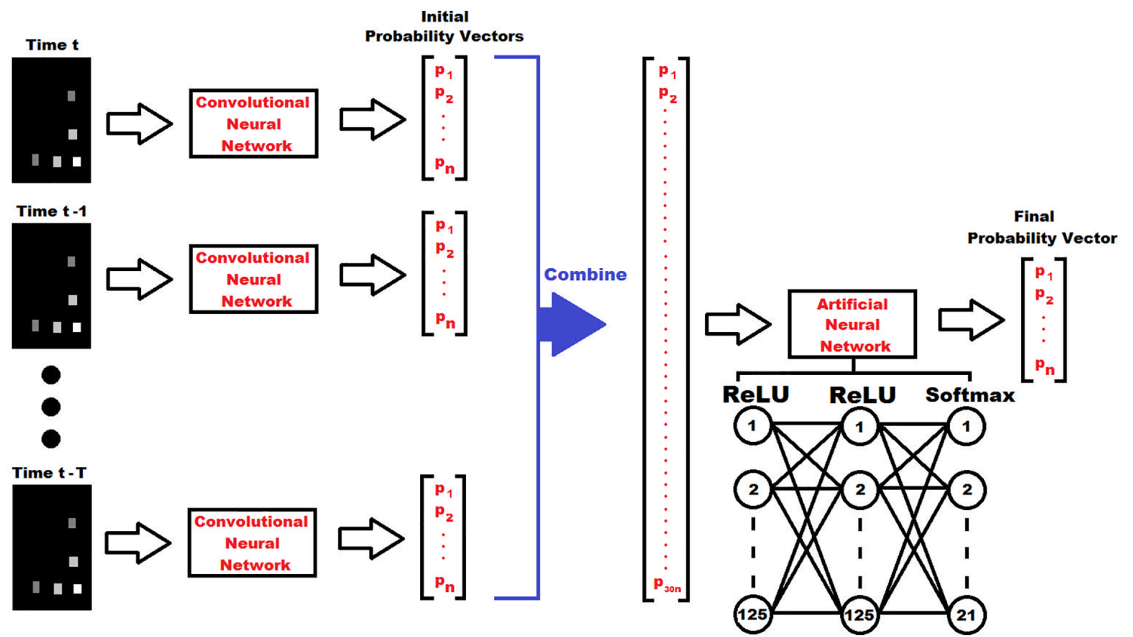

**Fig 4. Artificial neural network architecture.** The network structure for tracking the history of a tag is shown. First, grayscale images are made for each second of data until the desired history length, indicated by T, is reached. These images are then fed into the trained CNN shown in Fig 3. Next, the resulting probability vectors are combined together and go through an artificial neural network, yielding the final classification.

https://doi.org/10.1371/journal.pone.0205392.g004

### Neural network evaluation

The networks were evaluated using several important metrics, including precision, recall, $F_1$-scores, and accuracy. All of these metrics represent the rate of true positives (TP), false positives (FP), true negatives (TN), and false negatives (FN) in a myriad of ways:

$$Precision = \frac{TP}{TP + FP} \quad (2)$$

$$Recall = \frac{TP}{TP + FN} \quad (3)$$

$$F_1 score = \frac{2TP}{2TP + FP + FN} \quad (4)$$

$$Accuracy = \frac{TP + TN}{TP + TN + FP + FN} \quad (5)$$

Precision is also known as positive predictive value (PPV), whereas recall is also known as sensitivity. From the above equations, it is clear that as the number of false negatives and false positives approach zero, all of the scores approach one. Therefore, a perfect classifier will achieve a score of one for all of the above metrics.

In addition to evaluating the neural networks above, the data were also classified using more routine methods for comparison. The first method was RSSI thresholding, which classified each second of data based on the location of the maximum RSSI signal. Since some hallways were not equipped with any nodes (corridor halls 1-3), they were excluded from this classification and evaluation process. RSSI thresholding was also performed using thirty seconds of data by implementing majority voting. Essentially, each second of data was first classified as a particular zone, which counted as a vote for that zone. Then, after all thirty seconds were evaluated, the zone with the most votes would be the final classification. This thresholding and majority voting method (Threshold MV) was applied on the same data utilized by the ANN.

In addition, triangulation was also used to classify the location of the tag. First, the data were normalized and mapped onto a 34 x 15 pixel image (similar to above data processing) using an interpolation algorithm. In the data processing above, the points in between nodes were left as zeros. However, with the interpolation algorithm, all of the points in between were bicubicly interpolated based on all of the node signals. Then, the fifteen pixels with the highest values in this 34 x 15 map were identified. If these pixels belonged to different zones, the zone with the most pixels would be the final classification. Once again, the corridor halls were excluded from this method. Similar to the thresholding method, majority voting was also applied to the triangulation results in order to include temporal information, and these four methods (Thresholding, Thresholding MV, Triangulation, and Triangulation MV) were compared to the CNN and the CNN+ANN networks.

## Results

### Training results

The CNN training process took approximately 20 seconds per epoch for a total time of 29 minutes on a PC equipped with an Intel Core i7-4770 CPU clocked at 3.40GHz and 8 GB RAM. The training and validation losses are shown in Fig 5. Even after the tenth epoch, the training accuracy was already 90%, and continued to increase until all of the epochs were completed.

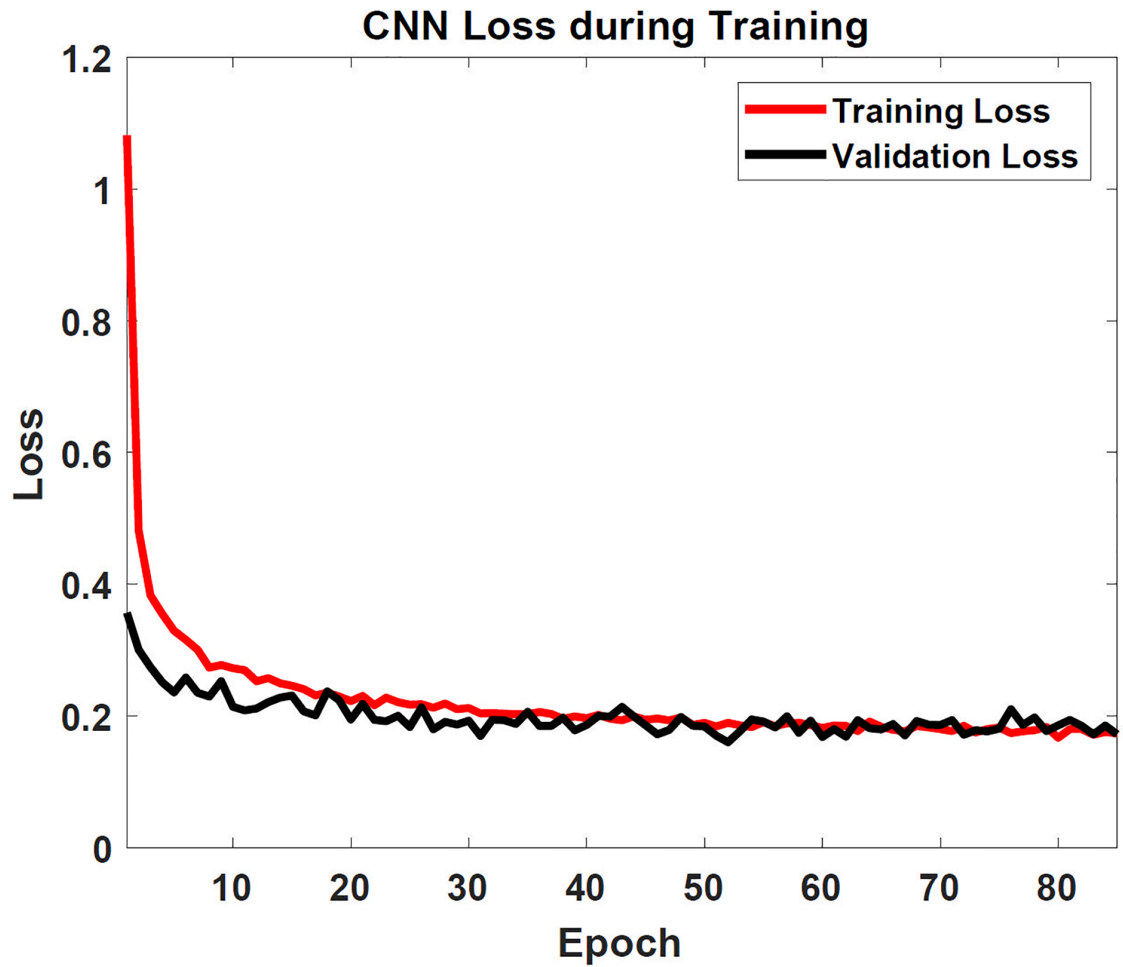

**Fig 5. Convolutional neural network loss function.** The categorical cross-entropy loss is shown for the convolutional neural network training as a function of the epoch number.

https://doi.org/10.1371/journal.pone.0205392.g005

From Fig 5, it is clear that over-fitting is not an issue for the CNN, since both the training and validation losses follow the same decreasing curve. This is further supported by the fact that the testing set had the exact same loss when it was evaluated after the CNN was trained. The ANN only took 30 seconds to train on the same PC, and yielded extraordinary results even after only one epoch: 98% accuracy. The loss quickly stabilized after the second epoch to a value close to zero, corresponding to an accuracy of 99.9%.

### Classification results

All of the classification methods (Thresholding, Thresholding MV, Triangulation, Triangulation MV, CNN, and CNN+ANN) were evaluated using the metrics described above, and the results can be seen in Fig 6 and Table 1. Since the most important rooms for proper clinical operation are the exam rooms, Fig 6 shows the results for exam rooms 1.1—1.12. In all cases, the CNN+ANN network outperforms all other methods and has a perfect score for all evaluation metrics. If temporal information is not used, the CNN network classifies the rooms much better than thresholding or triangulation. It is important to note that corridor halls 1-3 were

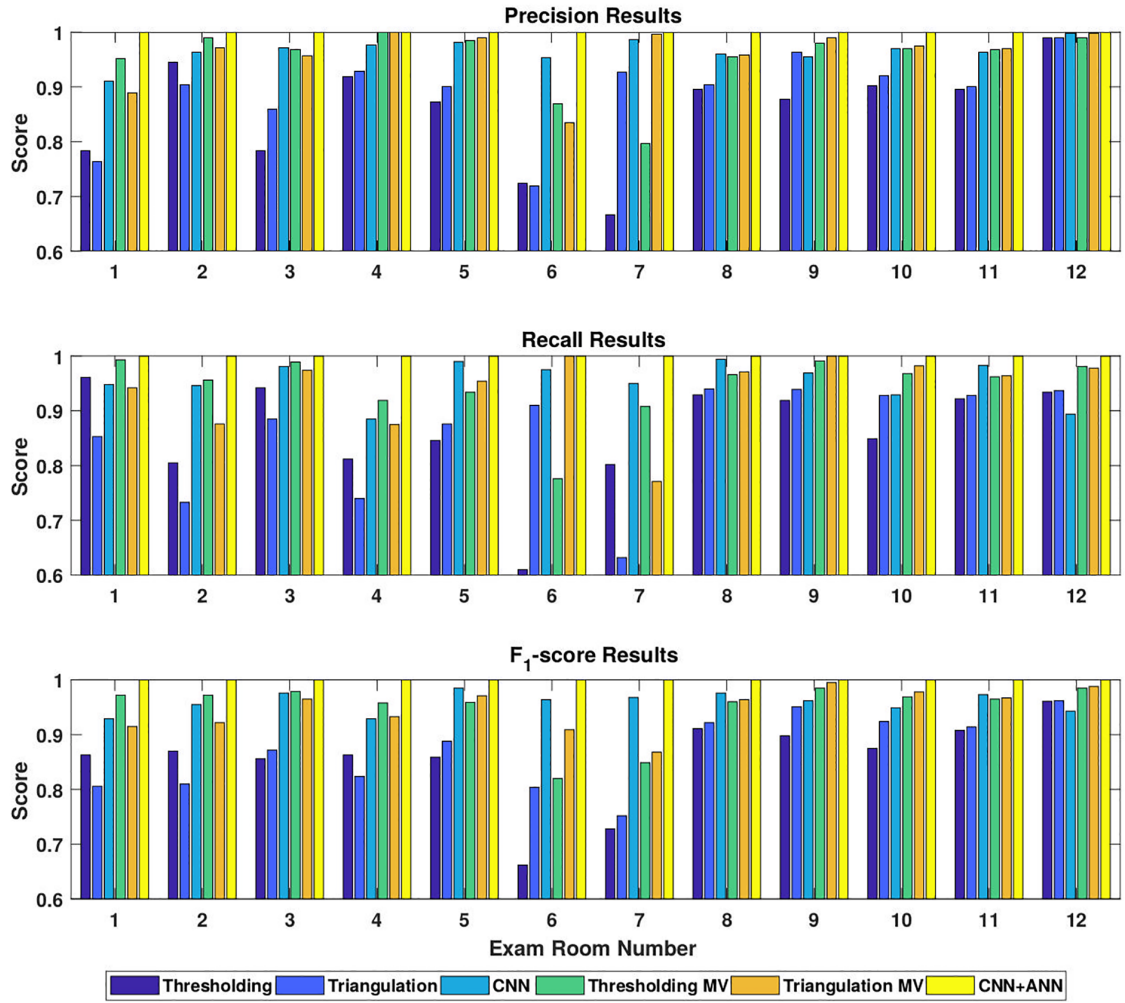

**Fig 6. Bar graph of classification results.** Precision, recall, and $F_1$-score results using the four classification methods (Thresholding, Thresholding with majority voting, CNN, and CNN+ANN) are shown.

https://doi.org/10.1371/journal.pone.0205392.g006

excluded from the thresholding and triangulation methods since these areas did not contain nodes, so including them would lower the performance of both techniques.

Table 1 shows the performance of the six methods over all applicable zones (18 zones for the thresholding and triangulation methods and 21 zones for neural networks). For this tabulation, the neural networks only evaluate the testing data when computing their scores. As seen from the table, it is clear that the CNN+ANN network performs perfectly. The CNN method

**Table 1. Classification results for each method.** The scores for the precision, recall, $F_1$-score, and accuracy metrics are shown for all four classification methods. The method with the highest score for each metric is highlighted in bold. The (†) symbol by the thresholding methods indicates that these methods only classified 18 zones, whereas the neural network methods classified 21 zones.

| Metric | Thresholding† | Triangulation† | CNN | Thresholding MV† | Triangulation MV† | CNN+ANN |
|---|---|---|---|---|---|---|
| Precision | 0.868 | 0.881 | 0.939 | 0.952 | 0.951 | **0.999** |
| Recall | 0.861 | 0.875 | 0.937 | 0.949 | 0.947 | **0.999** |
| $F_1$-Score | 0.861 | 0.874 | 0.937 | 0.949 | 0.946 | **0.999** |
| Accuracy | 0.861 | 0.875 | 0.937 | 0.949 | 0.947 | **0.999** |

https://doi.org/10.1371/journal.pone.0205392.t001

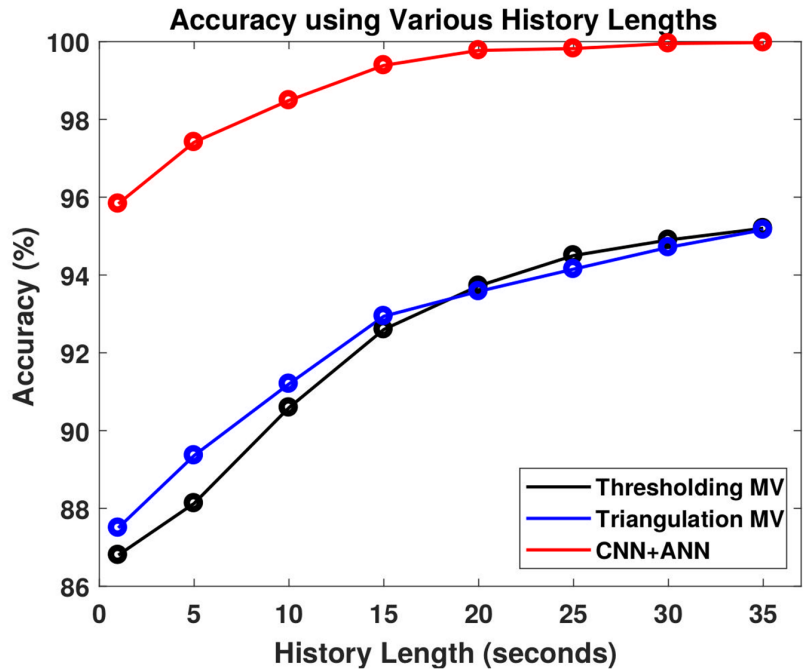

**Fig 7. Classification accuracy with history length.** The accuracy as a function of the history length (seconds) is displayed for the three classification methods: Thresholding with majority voting (Thresholding MV), Triangulation with majority voting (Triangulation MV), and the CNN+ANN network.

https://doi.org/10.1371/journal.pone.0205392.g007

performed better than both the thresholding and triangulation methods, even though the CNN classified 21 zones instead of only 18 zones. Adding temporal information through majority voting drastically improved the results of thresholding and triangulation, however the CNN+ANN performed better. Overall, it is clear from the results that adding temporal history to all of the classification methods helps to greatly improve precision, recall, $F_1$-score, and accuracy. Finally, Fig 7 demonstrates the affect of the history length on the accuracy for each classification method.

## Discussion

The main goal of this study was to assess the feasibility of classifying the location of BLE tags with neural networks. From Fig 6 and Table 1, it is clear that neural networks have clear advantages when compared to RSSI thresholding and triangulation techniques for determining the location of the tag. The thresholding method performed better than the triangulation method in this study due to the fact that the node placement and the tags were not optimized for triangulation. The BLE tags used were capable of sending signal but were not able to receive signal, which lowered the performance of the triangulation method. While artificial neural networks have previously been used for indoor BLE tracking [9], this study is the first to our knowledge that incorporates a convolutional neural network into the location tracking model. The temporal history of the tag was utilized by an ANN in this approach, but other methods such as a recurring neural network (RNN) [21, 22] could have also been used. Surprisingly, the CNN+ANN technique performed with 99.9% accuracy for the test set, which consisted of 30% of the total collected data. This is important because it is essential that changes made to the clinical workflow are based on accurate data.

Although the results indicate that neural networks are superior to classical methods for location tracking, it is important to note that there are limitations to the current study. The first major limitation is that all of the data were collected while the tags were in the same zone. In reality, a tag holder will be moving to different zones using the hallways, and it is very important to be able to follow the tag through these zone transitions. These transitions will also have adverse effects on the results for the methods that utilize the tag history, including thresholding and triangulation with majority voting, as well as the CNN+ANN network. However, by retraining the CNN+ANN network, it should be possible to restore accurate classification and retain the network architecture. In addition, the long short-term memory (LSTM) [22] network may be capable of learning better than the ANN used in this study to identify the transitions between each room. Another limitation of the current study is the limited amount of space used for data collection. The map shown in Fig 1A is only about 20% of the clinical area used in the radiation oncology clinic. The full clinical area contains Raspberry Pis in each important zone, and ideally the signals from all nodes in the facility should be used to determine tag location throughout the building. Both of these issues will be addressed in a future study.

Even with these current limitations, it is clear that the CNN+ANN network offers a great deal in terms of BLE tracking for the radiation oncology clinic. With further development, this system may become a viable alternative for clinics or businesses that cannot afford RFID technology. In addition, BLE technology is versatile and can be used in combination with a number of hardware and software for integration into the clinic. After this network has been tested more rigorously, the next hurdle will be to implement this system into everyday clinical use. Aside from collecting data on patient and staff traffic in the clinic, this system can have practical daily functions. For example, the system can notify relevant staff members if a patient has been left unattended in an exam room for a long period of time. With the versatility of deep learning methods, it is possible to eventually track other information, including biometric information, as well. Hopefully, these methods will lead to the development of a Smart Clinic, which will be able to use this information to improve the workflow and efficiency of the radiation oncology clinic.

## Conclusion

This study demonstrates the feasibility of the CNN+ANN approach for identifying the location of a BLE tag. The novel CNN+ANN method out-performed both RSSI thresholding and triangulation methods. This network will be further tested throughout the entire radiation oncology facility with the hope of utilizing this information to improve clinical workflow and patient safety in the future.

## Author Contributions

**Conceptualization:** Da Luo, Yulong Yan, Kenneth Westover, Hak Choy, Steve Jiang.

**Data curation:** Da Luo, Peter Henry.

**Formal analysis:** Zohaib Iqbal, Samaneh Kazemifar, Timothy Rozario.

**Funding acquisition:** Hak Choy, Steve Jiang.

**Investigation:** Da Luo, Peter Henry, Samaneh Kazemifar, Timothy Rozario, Yulong Yan, Kenneth Westover, Weiguo Lu, Dan Nguyen, Troy Long, Jing Wang, Steve Jiang.

**Methodology:** Zohaib Iqbal, Samaneh Kazemifar, Timothy Rozario, Jing Wang, Steve Jiang.

**Project administration:** Da Luo, Steve Jiang.

**Resources:** Hak Choy, Steve Jiang.

**Software:** Peter Henry, Yulong Yan.

**Supervision:** Yulong Yan, Kenneth Westover, Weiguo Lu, Jing Wang, Steve Jiang.

**Writing – original draft:** Zohaib Iqbal.

**Writing – review & editing:** Zohaib Iqbal, Da Luo, Dan Nguyen, Troy Long, Steve Jiang.